\documentclass[a4paper]{article}

\usepackage{INTERSPEECH2019}
\usepackage{xcolor}
\usepackage{amssymb}
\usepackage{pifont}

\title{Tongji University Team for the VoxCeleb Speaker Recognition Challenge 2020}
\name{Rui Wang$^1$, Zhihua Wei$^1$, Yibin Zhan$^1$, Zhuoxi Chen$^1$}
\address{
  $^1$Department of Computer Science and Technology, Tongji University, Shanghai, China}
\email{\{rwang,zhihua\_wei,1633044,oyugu\}@tongji.edu.cn}

\begin{document}

\maketitle
\begin{abstract}
In this report, we describe the submission of Tongji University team to the CLOSE track of the VoxCeleb Speaker Recognition Challenge (VoxSRC) 2020 at Interspeech 2020. We investigate different speaker recognition systems based on the popular ResNet-34 architecture, and train multiple variants via various loss functions. Both Offline and online data augmentation are introduced to improve the diversity of the training set, and score normalization with the exhaustive grid search is applied in the post-processing. Our best fusion of five selected systems for the CLOSE track achieves 0.2800 DCF and 4.7770\% EER on the challenge.
\end{abstract}

\section{Introduction}

The VoxCeleb Speaker Recognition Challenge (VoxSRC) 2020 is second installment of the new series of speaker recognition challenges that are hosted annually. The challenge is intended to assess how well current speaker recognition technology enables to identify speakers in unconstrained or 'in the wild' condition. This year's challenge is different to the last in two aspects: (1) there is an explicit domain shift between the training data and the test data \cite{Heo2020}; (2) the evaluation set contains more than half of utterances that ranges 2 seconds to 4 seconds.

Deep neural network-based approaches have been proven to be helpful to embedding extraction in speaker recognition \cite{chung2020in, Snyder2017b}. All of our systems to embedding extraction use well-known ResNet-34 topology \cite{he2016deep}. The architecture using residual connections in the block can be helpful to its training, and it has achieved promising performance in various tasks.

The rest of this report is organized as follows: in Section~\ref{sec:model}, we describe the components of our model, such as input representation, network architecture, as well as loss function. In Section~\ref{sec:experiments}, the setup of experiments are presented. Subsequently, the results and analysis are given in Section~\ref{sec:results}. Finally, we conclude the report in Section~\ref{sec:conclusion}.

\section{Model}\label{sec:model}

\subsection{Input feature}

During training, we use a 2-second (32000 samples in 16 kHz) segment extracted randomly from each utterance, but use a longer segment in the evaluation, e.g., 4 seconds. The shorter utterance is padded to the minimum duration in the wrap mode as \cite{Zhang2017}. No voice activity detection is applied due to most of the continuous speech in the VoxCeleb dataset.

There are two different features as follows:

\begin{itemize}
\item \textbf{40-dim log Mel-filterbanks (40 FB).} Spectrograms are extracted with a hann window of width 25ms and step 10ms with a fast fourier transform (FFT) size of 512. Mel-filterbanks are then extracted from spectrogram within the given frequency limit 20-7600 Hz.
\item \textbf{64-dim log Mel-filterbanks (64 FB).} Pre-emphasis is first applied to the input signal using a coefficient of 0.97. Spectrograms are extracted with a hamming window of width 25ms and step 10ms with a FFT size of 512.
\end{itemize}

Mean and variance normalization is applied to both features by using instance normalization.

\subsection{Trunk architecture}

\begin{figure}[t]
  \centering
  \includegraphics[width=5.4cm]{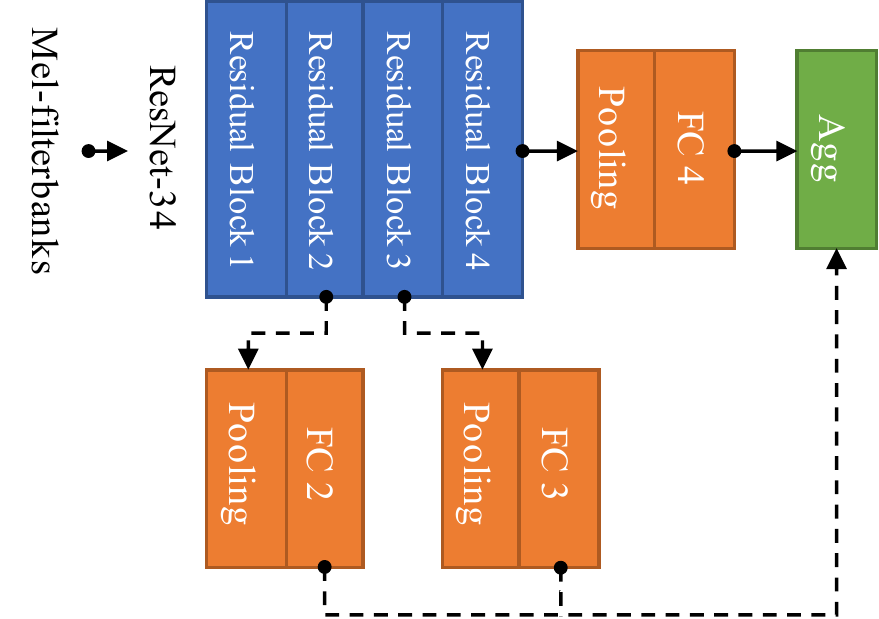}
  \caption{An overview of multiple-embedding aggregator. Directed lines are connection between components, where the dash ones denote alternative connection and are different from the solid ones.}
  \label{fig:arch}
\end{figure}

We propose two variants of the ResNet with 34 layers:\\\\
\textbf{Half-channels ResNet-34.} This architecture has half of the channels in each residual block compared to the ResNet-34 and contains over 5.8 million parameters. The stride at the first convolutional layer is also removed that leads to increased computational requirements. Temporal pooling layer, such as statistics pooling (SP) or attentive statistics pooling (ASP) \cite{okabe2018attentive}, is used to aggregate temporal frames. Table~\ref{tab:arch} shows the detailed architecture of the model. We refer to this configuration as \textbf{H-SP} or \textbf{H-ASP} corresponding to the ResNet-34 with SP or with ASP, respectively.
\begin{table}[h]
  \caption{Trunk architecture for the half-channels ResNet-34. The shape of input is $\textbf{\emph{L}}\times\textbf{\emph{D}}\times1$. \textbf{\emph{L:}} the length of the input sequence. \textbf{\emph{D:}} the dimension of the feature, e.g., 40 and 64. \textbf{\emph{Pool:}} temporal pooling layers, e.g., statistics pooling (SP) and attentive statistics pooling (ASP). \textbf{\emph{M:}} the size of embedding extracted from the trunk architecture.}
  \label{tab:arch}
  \centering
  \begin{tabular}{lccr}
  	\toprule
	\textbf{Layer} & \textbf{Kernel size} & \textbf{Stride} & \textbf{Output shape} \\
  	\midrule
	Conv1 & $3\times3\times32$ & $1\times1$ & $32\times D\times L$\\
	\midrule
	Res1 & $3\times3\times32$ & $1\times1$ & $32\times D\times L$\\
	\midrule
	Res2 & $3\times3\times64$ & $2\times2$ & $64\times \frac{D}{2}\times L$\\
	\midrule
	Res3 & $3\times3\times128$ & $2\times2$ & $128\times \frac{D}{4}\times L$\\
	\midrule
	Res4 & $3\times3\times256$ & $2\times2$ & $256\times \frac{D}{8}\times L$\\
	\midrule
	Pool & - & - & $\frac{D}{4}\times256$\\
	\midrule
	Flatten & - & - & $64D$\\
	\midrule
	Linear & $M$ & - & $M$\\
	\bottomrule
  \end{tabular}
\end{table}\\
\textbf{Multiple-Embedding Aggregator.} This architecture also has half of the channels in each residual block. In addition, inspired by \cite{Hajavi2019}, multiple embeddings that come from former outputs of residual blocks and pass through the identical pooling layer are aggregated by vector-wise weighting. Figure~\ref{fig:arch} give an overview of the architecture. We refer to this configuration as \textbf{H2} or \textbf{H3} corresponding to the model that aggregates 2 or 3 outputs of residual blocks.

\subsection{Loss function}

We train the networks using three types of loss functions, e.g., softmax loss, additive angular margin softmax loss (AAM-Softmax) \cite{deng2019arcface}, as well as angular prototypical loss \cite{chung2020in}. 
\begin{itemize}
\item \textbf{Softmax loss (S)} is a common classification-based loss.
\item \textbf{Additive angular margin softmax loss (AAM)} introduces a concept of a margin between classes where the margin increases inter-class variance. We use a margin of 0.2 and a scale of 30 as \cite{chung2020in}.
\item \textbf{Angular prototypical loss (AP)} is a variant of the prototypical networks with an angular objective. It results in multiple times of memory requirements as the increased total number of the support set and query set. We use the total number of 2.
\end{itemize}

Two combination of the loss functions are used to training: 

\begin{itemize}
\item \textbf{The angular prototypical loss with additional of the softmax loss (AP+S)}. Both losses are summed.
\item \textbf{The softmax loss followed by the additive angular margin softmax loss (S-AAM)}. The softmax loss function is applied to train a model with several epochs, then the additive angular margin softmax loss is applied.
\end{itemize}

\section{Experiments}\label{sec:experiments}

\subsection{Dataset}

We submit our results to the CLOSE track. This track allow only the development set of VoxCeleb2 \cite{Chung2018}, which contains 5,994 speakers. For data augmentation, we use the MUSAN corpus \cite{snyder2015musan} and RIR corpus \cite{ko2017study} to generate augmented sets. For evaluation, we use the VoxCeleb1 test sets \cite{Nagrani2017} and the VoxSRC 2020 validation set.

\subsection{Data augmentation}

We use two augmentation methods in speech processing, e.g., additive noise and room impulse response (RIR) simulation. For additive noise, we use the MUSAN corpus, which contains three types of voice, such as human speech, music, and background noises. Specifically, these utterances are segmented into 5-second width and 3-second step in consideration of the speed of reading a segment during training. For RIRs, we apply the simulated RIR filters in the simplified version of \cite{ko2017study}. Both noise and RIR filters are randomly selected in a training step.

We perform data augmentation with two settings: offline and online. In the offline settings, music and noise are used as additional noises, RIRs are performed from small, medium and large rooms. Also, time mask and frequency mask are introduced as \cite{Park2019}. It totally creates $5\times$ the original VoxCeleb2 training set. The online settings, similar to \cite{Snyder2018}, applies the four methods as mentioned in \cite{Heo2020}, in which the recordings are augmented by one of the following methods.

\begin{itemize}
\item \textbf{Speech}: 3 to 7 recordings are randomly picked from MUSAN, then added to the original input signal with a random signal to noise ratio (SNR) from 13 to 20dB.
\item \textbf{Music}: A single music utterances is randomly picked from MUSAN, then added to the original input signal from 5 to 15dB SNR.
\item \textbf{Noise}: A background noises in MUSAN is randomly picked, then added to the recording from 0 to 15dB SNR.
\item \textbf{RIR filters}: Reverberation is performed by convolution operation with a RIR filters, which is normalized via the power of the recording.
\end{itemize}

\subsection{Implementation details}

In our implementation, the trunk architecture are realized by the PyTorch framework \cite{Paszke2019}; the score normalization and the system fusion are realized by various pythonic tools. The models are trained using a single NVIDIA V100 GPU with 32GB memory with the Adam optimizer. One epoch is defined as a full pass through the training dataset, and the size of the training set in the offline data augmentation is as 5 times as that in online or no settings.

\begin{itemize}
\item \textbf{No data augmentation (No)}. We use an initial learning rate of 0.001, reduced by 5\% every 5 epochs. The softmax is first used to train the network for 30 epochs, and then the pre-trained model is trained by AAM-Softmax for 200 epochs. We use batch size of 200. The maximum number of utterances per speaker is 200. The training process takes around 4 days to train.
\item \textbf{Offline data augmentation (Off)}. We use an initial learning rate of 0.001, reduced by 10\% or 5\% every epoch. The softmax is first used to train the network for 3 epochs. Then the AAM-Softmax is applied to train the network for 32 epochs or less. We use batch size of 128. The maximum number of utterances per speaker is 2000 or 2500. The training process takes 5-10 days to train.
\item \textbf{Online data augmentation (On)}. We use an initial learning rate of 0.001, reduced by 25\% every 24 epochs. The network is trained for 360 epochs or more. A weight decay of 5e-5 is applied. We use batch size of 200. The process is  as a single-GPU version as \cite{Heo2020}. The models take over 10 days to train.
\end{itemize}

\subsection{Cosine Scoring}

The trained networks are evaluated in three VoxCeleb1 test sets and the VoxSRC 2020 validation set. We sample 10 segments of 4 seconds at regular intervals from each test segment in a uniform step, and compute the 100 ($10\times10$) cosine similarities between the possible combinations from every pair of segments. The mean of the 100 similarities is used as the score. This protocol is in line with those works in \cite{Heo2020, chung2020in, Chung2018, Chung2020}.

\begin{table}[th]
  \caption{The results of score normalization based on clova baseline on the VoxSRC 2020 validation set. All experiments are repeated 10 times. We report the mean and standard deviation.}
  \label{tab:score_normalization}
  \centering
  \begin{tabular}{ccrr}
  	\toprule
	\textbf{N} & \textbf{X} & \textbf{EER} & \textbf{DCF} \\
  	\midrule
	2000 & 200 & $3.6246 \pm 0.0460$ & $0.2032 \pm 0.0042$ \\
	3000 & 200 & $\mathbf{3.5867 \pm 0.0274}$ & $0.1975 \pm 0.0022$ \\
	4000 & 200 & $3.5904 \pm 0.0328$ & $\mathbf{0.1947 \pm 0.0027}$ \\
	2000 & 300 & $3.6686 \pm 0.0382$ & $0.2086 \pm 0.0034$ \\
	3000 & 300 & $3.5995 \pm 0.0352$ & $0.2004 \pm 0.0027$ \\
	4000 & 300 & $3.5990 \pm 0.0307$ & $0.1982 \pm 0.0025$ \\
	2000 & 400 & $3.6879 \pm 0.0588$ & $0.2113 \pm 0.0052$ \\
	3000 & 400 & $3.6336 \pm 0.0364$ & $0.2043 \pm 0.0042$ \\
	4000 & 400 & $3.6285 \pm 0.0201$ & $0.2016 \pm 0.0023$ \\
	2000 & 2000 & $4.1799 \pm 0.0738$ & $0.2566 \pm 0.0070$ \\
	3000 & 3000 & $4.1119 \pm 0.0405$ & $0.2508 \pm 0.0040$ \\
	4000 & 4000 & $4.1329 \pm 0.0330$ & $0.2532 \pm 0.0027$ \\
	\bottomrule
  \end{tabular}
\end{table}

\begin{table*}[th]
  \caption{Results on the VoxCeleb test sets and VoxSRC validation set. \textbf{\emph{H/ASP-512:}} H/ASP model with 512-dimensional embedding. \textbf{\emph{FB:}} log mel-scale Filter Banks. \textbf{\emph{S:}}. Softmax. \textbf{\emph{AAM:}} Additive Angular Margin Softmax. \textbf{\emph{AP:}} Angular Prototypical. \textbf{\emph{SN:}} adaptive Symmetric score Normalisation. \textbf{\emph{EER:}} Equal Error Rate (\%). \textbf{\emph{DCF:}} Minimum Detection Cost of the Function.}
  \label{tab:results}
  \centering
  \begin{tabular}{rlllll|cc|cc|cc|cccc}
  	\toprule
	\textbf{Sys.} & \textbf{Config.} & \textbf{Loss} & \textbf{FB} & \textbf{Aug.} & \textbf{SN} & \multicolumn{2}{|c}{\textbf{Vox1}} & \multicolumn{2}{|c}{\textbf{Vox1-E cl.}} & \multicolumn{2}{|c}{\textbf{Vox1-H cl.}} & \multicolumn{2}{|c}{\textbf{SRC 2020 Val.}}\\
	& & & & & & \textbf{EER} & \textbf{DCF} & \textbf{EER} & \textbf{DCF} & \textbf{EER} & \textbf{DCF} & \textbf{EER} & \textbf{DCF}\\
  	\midrule
	- & FR-34 \cite{chung2020in} & AP & - & No & \ding{51} & 2.22 & - & - & - & - & - & - & -\\
	- & BUT \cite{Zeinali2019} & S-AAM & 40 & Off & \ding{51} & - & - & 1.35 & - & 2.48 & - & - & -\\
	- & Fusion \cite{Zeinali2019} & - & - & - & - & - & - & 1.14 & - & 2.21 & - & - & -\\
	- & H/ASP \cite{Heo2020} & AP+S & 64 & On & \ding{55} & 1.18 & 0.086 & 1.21 & 0.086 & 2.38 & 0.154 & 3.79 & 0.213\\
	\midrule
	1 & H/ASP-512 & - & - & - & \ding{51} & \textbf{1.12} & \textbf{0.088} & \textbf{1.13} & \textbf{0.078} & \textbf{2.22} & \textbf{0.136} & \textbf{3.55} & \textbf{0.190} \\ 
	2 & H/ASP-512 & AP+S & 64 & No & \ding{55} & 1.88 & 0.141 & 1.86 & 0.132 & 3.37 & 0.212 & 5.00 & 0.299 \\ 
	3 & H/SP-160 & S-AAM & 40 & No & \ding{55} & 2.33 & 0.154 & 2.22 & 0.149 & 3.79 & 0.226 & 5.53 & 0.309 \\ 
	4 & H/SP-160 & S-AAM & 40 & Off & \ding{55} & 1.59 & 0.114 & 1.74 & 0.117 & 3.13 & 0.193 & 4.66 & 0.266 \\ 
	5 & H/SP-160 & S-AAM & 40 & Off & \ding{51} & 1.51 & 0.118 & 1.68 & 0.119 & 3.00 & 0.191 & \textbf{4.38} & 0.258 \\ 
	6 & H/SP-256 & S-AAM & 40 & No & \ding{55} & 2.30 & 0.164 & 2.23 & 0.150 & 3.95 & 0.241 & 5.63 & 0.324 \\ 
	7 & H/SP-256 & S-AAM & 40 & Off & \ding{55} & \textbf{1.46} & \textbf{0.105} & 1.69 & \textbf{0.111} & 3.03 & 0.188 & 4.67 & 0.257 \\ 
	8 & H/SP-256 & S-AAM & 40 & Off & \ding{51} & 1.55 & 0.111 & \textbf{1.64} & 0.115 & \textbf{2.90} & \textbf{0.187} & \textbf{4.38} & \textbf{0.251} \\ 
	9 & H2/SP-256 & S-AAM & 40 & No & \ding{55} & 1.86 & 0.128 & 2.02 & 0.134 & 3.52 & 0.211 & 5.18 & 0.287 \\ 
	10 & H2/SP-256 & S-AAM & 40 & Off & \ding{55} & 1.55 & 0.117 & 1.68 & 0.113 & 3.15 & 0.195 & 4.68 & 0.266 \\ 
	11 & H2/SP-256 & S-AAM & 40 & Off & \ding{51} & 1.62 & 0.123 & 1.66 & 0.116 & 3.03 & 0.195 & 4.44 & 0.265 \\ 
	12 & H3/SP-256 & S-AAM & 40 & No & \ding{55} & 1.94 & 0.131 & 2.11 & 0.140 & 3.63 & 0.218 & 5.28 & 0.294 \\ 
	13 & H3/SP-256 & S-AAM & 40 & Off & \ding{55} & 1.69 & 0.117 & 1.68 & 0.113 & 3.20 & 0.197 & 4.78 & 0.266 \\ 
	14 & H3/SP-256 & S-AAM & 40 & Off & \ding{51} & 1.51 & 0.122 & 1.68 & 0.120 & 3.19 & 0.203 & 4.56 & 0.264 \\ 
	\midrule
	\multicolumn{6}{c|}{Sys 4 + Sys 7 + Sys 10 + Sys 13} & 1.33 & 0.092 & 1.49 & 0.098 & 2.76 & 0.169 & 4.25 & 0.245\\
	\multicolumn{6}{c|}{.9 Sys 1 + .05 Sys 4 + .01 Sys 7 + .03 Sys 10 + .01 Sys 13} & 1.09 & 0.084 & 1.11 & 0.077 & 2.18 & 0.133 & 3.50 & 0.187\\
	\multicolumn{6}{c|}{.7 Sys 1 + .08 Sys 5 + .07 Sys 8 + .06 Sys 11 + .09 Sys 14} & 1.07 & 0.080 & 1.10 & 0.077 & 2.14 & 0.131 & 3.37 & 0.186\\
	\bottomrule
  \end{tabular}
\end{table*}

\subsection{Score normalization}

For the cosine scoring, we used adaptive symmetric score normalization (AS-norm) which computes an average of normalized scores from Z-norm and T-norm \cite{Matejka2017, Sturim2005}. Firstly, the vectors of $N$ utterances are sampled randomly as the cohort set from the development set of VoxCeleb2. Next, $X$ top scoring or most similar utterances in the cohort are selected to compute mean and variance for normalization. The grid search is applied to achieve the process of selection. Finally, the cohort with the best performance is determined. Table~\ref{tab:score_normalization} shows the results, and $N=400$ and $X=200$ is selected due to the best DCF.

\subsection{System fusion}

For system fusion, we applied the approach similar to \cite{Variani2014}, which sums the scores provided by individual system for each trial, and the score normalization is used in every system to facilitate the combination of scores. Specifically, firstly, we select the trained model with the optimal performance in each architecture as candidates. Secondly, these systems are fused in consideration of the performance of separate systems, where the system with a smaller DCF picks a higher weight. Since the scores of different systems vary, we normalize scores into the closed interval $[0, 1]$ via min-max scale before fusion. Also, the sum of weights is limited to 1. Finally, we obtain five weights by using the trial-and-error mean in terms of DCF on the VoxSRC 2020 validation set.

\subsection{Evaluation protocol}

We report two performance metrics: (1) the Equal Error Rate (EER \%) which is the rate at which both acceptance ($E_\text{fa}$) and rejection ($E_\text{miss}$) errors are equal or very approximate; (2) the minimum detection cost of the function (DCF) used by the NIST SRE. In particular, the parameters $C_\text{miss} = 1$, $C_\text{fa} = 1$ and $P_\text{target} = 0.05$ are used for the cost function as follows:

\begin{equation}
\text{DCF} = C_\text{miss}\cdot E_\text{miss}\cdot P_\text{target} + C_\text{fa}\cdot E_\text{fa}\cdot (1-P_\text{target})
\end{equation}

\section{Results and discussion}\label{sec:results}

Table~\ref{tab:results} reports the experimental results. The clova baseline system remains the optimal performance in the case of no use of model ensemble or post-processing. 

\subsection{Analysis of score normalization}

We apply AS-norm to the clova pre-trained models \cite{Heo2020} and achieves improvement by around 10\%. However, AS-norm does not work well on other models and obains sight or even little gain. One probable explanation is that the model trained by angular prototypical loss can benefit the score normalization compared to the models trained by AAM-Softmax loss.

\subsection{Analysis of data augmentation}

For no data augmentation, the multiple-embedding aggregator is the optimal, and H2/SP-256 can achieve 0.287 DCF on the VoxSRC 2020 validation set. For offline augmentation, the ResNet-like architectures without embeddings aggregation can obtain better performance, e.g., 0.256 DCF in H/SP-256. Also, these models without embeddings aggregation is more suitable to be score normalized than the model with the aggregator. On the other hand, online data augmentation provides around 10\% gain compared to the offline settings. For example, the model with VoxCeleb1 1.34\% EER trained via online data augmentation can produce 4.08\% EER on the VoxSRC 2020 validation set (we do not present this result in the report.), but the model with VoxCeleb1 1.46\% EER trained via offline data augmentation produces 4.67\% EER.

\subsection{Analysis of system fusion}

For system fusion, it is to note that sight improvement is obtained in case the fusion of systems are not comparable in each other, e.g., 3.55 DCF $\to$ 3.37 DCF on the VoxSRC 2020 validation set. However, a significant gain can be achieved by fusing the models with comparable performance, e.g., 4.66 DCF $\to$ 4.25 DCF on the VoxSRC 2020 validation set.

\subsection{Results of submission}

For the CLOSE track, the results of submission are as follows:

\begin{itemize}
\item \textbf{H/ASP-512 with AS-norm}. The system with score normalization achieves 0.2980 DCF and 5.1670\% EER.
\item \textbf{Fusion of System 1, 4, 7, 10 and 13}. The fusion system achieves 0.2910 DCF and 5.0730\%.
\item \textbf{Fusion of System 1, 5, 8, 11 and 14}. The system fusing score-normalized systems further improves and achieves 0.2800 DCF and 4.7770\%.
\end{itemize}

\section{Conclusions}\label{sec:conclusion}

The report describes the submission of Tongji University team for the 2020 VoxSRC Speaker Recognition Challenge. We give a comprehensive analysis on the ResNet-34 architecture, score normalization, as well as system fusion. Our best fusion outperforms one ensemble system in the last year’s challenge and achieves 0.2800 DCF and 4.7770\%. In the future, we are going to consider speaker recognition in the open training datasets and self-supervised condition.

\section{Acknowledgements}

This work is partially supported by the National Key Research and Development Project (No. 213), the National Nature Science Foundation of China (No. 61976160).


\bibliographystyle{IEEEtran}

\bibliography{mybib}

\begin{thebibliography}{10}
\providecommand{\url}[1]{#1}
\csname url@samestyle\endcsname
\providecommand{\newblock}{\relax}
\providecommand{\bibinfo}[2]{#2}
\providecommand{\BIBentrySTDinterwordspacing}{\spaceskip=0pt\relax}
\providecommand{\BIBentryALTinterwordstretchfactor}{4}
\providecommand{\BIBentryALTinterwordspacing}{\spaceskip=\fontdimen2\font plus
\BIBentryALTinterwordstretchfactor\fontdimen3\font minus
  \fontdimen4\font\relax}
\providecommand{\BIBforeignlanguage}[2]{{%
\expandafter\ifx\csname l@#1\endcsname\relax
\typeout{** WARNING: IEEEtran.bst: No hyphenation pattern has been}%
\typeout{** loaded for the language `#1'. Using the pattern for}%
\typeout{** the default language instead.}%
\else
\language=\csname l@#1\endcsname
\fi
#2}}
\providecommand{\BIBdecl}{\relax}
\BIBdecl

\bibitem{Heo2020}
H.~S. Heo, B.-J. Lee, J.~Huh, and J.~S. Chung, ``Clova baseline system for the
  voxceleb speaker recognition challenge 2020,'' \emph{arXiv preprint
  arXiv:2009.14153}, pp. 1--3, 2020.

\bibitem{chung2020in}
J.~S. Chung, J.~Huh, S.~Mun, M.~Lee, H.~S. Heo, S.~Choe, C.~Ham, S.~Jung, B.-J.
  Lee, and I.~Han, ``In defence of metric learning for speaker recognition,''
  in \emph{Interspeech}, 2020.

\bibitem{Snyder2017b}
D.~Snyder, D.~Garcia-Romero, D.~Povey, and S.~Khudanpur, ``Deep neural network
  embeddings for text-independent speaker verification,'' in
  \emph{Interspeech}, 2017, pp. 999--1003.

\bibitem{he2016deep}
K.~He, X.~Zhang, S.~Ren, and J.~Sun, ``Deep residual learning for image
  recognition,'' in \emph{IEEE Conference on Computer Vision and Pattern
  Recognition}, 2016, pp. 770--778.

\bibitem{Zhang2017}
C.~Zhang and K.~Koishida, ``End-to-end text-independent speaker verification
  with triplet loss on short utterances,'' in \emph{Interspeech}, 2017, pp.
  1487--1491.

\bibitem{okabe2018attentive}
K.~Okabe, T.~Koshinaka, and K.~Shinoda, ``Attentive statistics pooling for deep
  speaker embedding,'' in \emph{Interspeech}, 2018, pp. 2252--2256.

\bibitem{Hajavi2019}
A.~Hajavi and A.~Etemad, ``A deep neural network for short-segment speaker
  recognition,'' in \emph{Interspeech}, 2019, pp. 2878--2882.

\bibitem{deng2019arcface}
J.~Deng, J.~Guo, N.~Xue, and S.~Zafeiriou, ``Arcface: Additive angular margin
  loss for deep face recognition,'' in \emph{IEEE Conference on Computer Vision
  and Pattern Recognition}, 2019, pp. 4690--4699.

\bibitem{Chung2018}
J.~S. Chung, A.~Nagrani, A.~Zisserman, and A.~{Int Speech Commun},
  ``{VoxCeleb2}: {Deep} speaker recognition,'' in \emph{Interspeech}, 2018, pp.
  1086--1090.

\bibitem{snyder2015musan}
D.~Snyder, G.~Chen, and D.~Povey, ``Musan: A music, speech, and noise corpus,''
  \emph{arXiv preprint arXiv:1510.08484}, 2015.

\bibitem{ko2017study}
T.~Ko, V.~Peddinti, D.~Povey, M.~L. Seltzer, and S.~Khudanpur, ``A study on
  data augmentation of reverberant speech for robust speech recognition,'' in
  \emph{ICASSP}, 2017, pp. 5220--5224.

\bibitem{Nagrani2017}
A.~Nagrani, J.~S. Chung, and A.~Zisserman, ``{VoxCeleb}: {A} large-scale
  speaker identification dataset,'' in \emph{Interspeech}, 2017, pp.
  2616--2620.

\bibitem{Park2019}
D.~S. Park, W.~Chan, Y.~Zhang, C.~C. Chiu, B.~Zoph, E.~D. Cubuk, and Q.~V. Le,
  ``{Specaugment}: {A} simple data augmentation method for automatic speech
  recognition,'' in \emph{Interspeech}, 2019, pp. 2613--2617.

\bibitem{Snyder2018}
D.~Snyder, D.~Garcia-Romero, G.~Sell, D.~Povey, and S.~Khudanpur,
  ``{X-Vectors}: {Robust DNN} embeddings for speaker recognition,'' in
  \emph{ICASSP}, 2018, pp. 5329--5333.

\bibitem{Paszke2019}
A.~Paszke, S.~Gross, F.~Massa, A.~Lerer, J.~Bradbury, G.~Chanan, T.~Killeen,
  Z.~Lin, N.~Gimelshein, L.~Antiga, A.~Desmaison, A.~K{\"{o}}pf, E.~Yang,
  Z.~DeVito, M.~Raison, A.~Tejani, S.~Chilamkurthy, B.~Steiner, L.~Fang,
  J.~Bai, and S.~Chintala, ``{PyTorch: An} imperative style, high-performance
  deep learning library,'' in \emph{NIPS}, 2019, pp. 8024--8035.

\bibitem{Chung2020}
J.~S. Chung, J.~Huh, and S.~Mun, ``{Delving into VoxCeleb: Environment}
  invariant speaker recognition,'' in \emph{Odyssey}, 2020, pp. 349--356.

\bibitem{Zeinali2019}
H.~Zeinali, S.~Wang, A.~Silnova, P.~Mat{\v{e}}jka, and O.~Plchot, ``{BUT}
  system description to {VoxCeleb} speaker recognition challenge 2019,''
  \emph{arXiv preprint arXiv:1910.12592}, pp. 4--7, 2019.

\bibitem{Matejka2017}
P.~Mat{\v{e}}jka, O.~Novotn{\'{y}}, O.~Plchot, L.~Burget, M.~D. S{\'{a}}nchez,
  and J.~{\v{C}}ernock{\'{y}}, ``Analysis of score normalization in
  multilingual speaker recognition,'' in \emph{Interspeech}, 2017, pp.
  1567--1571.

\bibitem{Sturim2005}
D.~Sturim and D.~Reynolds, ``Speaker adaptive cohort selection for tnorm in
  text-independent speaker verification,'' in \emph{ICASSP}, 2005, pp.
  741--744.

\bibitem{Variani2014}
E.~Variani, X.~Lei, E.~McDermott, I.~L. Moreno, and J.~Gonzalez-Dominguez,
  ``Deep neural networks for small footprint text-dependent speaker
  verification,'' in \emph{ICASSP}, 2014, pp. 4052--4056.

\end{thebibliography}

\end{document}